\documentclass{emulateapj} 

\usepackage{epsfig}
\newcommand{\unit}[1]{\ifmmode {\rm\ #1} \else {$\rm #1$} \fi}

\newcommand{\AlII}{{\rm Al\,{\sc ii}}}

\newcommand{\CI}{{\rm C\,{\sc i}}}
\newcommand{\CIII}{{\rm C\,{\sc iii}}}
\newcommand{\CIV}{{\rm C\,{\sc iv}}}

\newcommand{\Htwo}{{\rm \unit{H_2}}}
\newcommand{\HeII}{{\rm He\,{\sc ii}}}

\newcommand{\OIIIb}{{\rm O\,{\sc iii]}}}
\newcommand{\OIVb}{{\rm O\,{\sc iv]}}}
\newcommand{\OI}{{\rm O\,{\sc i}}}
\newcommand{\OVI}{{\rm O\,{\sc vi}}}

\newcommand{\NII}{{\rm N\,{\sc ii}}}
\newcommand{\NIVb}{{\rm N\,{\sc iv]}}}
\newcommand{\SiII}{{\rm Si\,{\sc ii}}}
\newcommand{\SiIIs}{\SiII$^{*}$}
\newcommand{\SiIV}{{\rm Si\,{\sc iv}}}

\newcommand{\angstrom}{\unit{\AA}}

\newcommand{\lu}{\unit{ph\ s^{-1}}\-\unit{cm^{-2}}\- \unit{sr^{-1}}}

\newcommand{\degrees}{\ifmmode {^{\circ}} \else {$^{\circ}$}\fi}
\newcommand{\degree}{\ifmmode {^{\circ}} \else {$^{\circ}$}\fi}

\newcommand{\doublet}{\ifmmode {\lambda\lambda} \else {$\lambda\lambda$} \fi}

\newcommand{\quarter}{\ifmmode {\frac{1}{4}} \else {$\frac{1}{4}$} \fi}

\newcommand{\singlet}{\ifmmode {\lambda} \else {$\lambda$} \fi}

\newcommand{\tentothe}[1]{\ifmmode {10^{#1}} \else {$10^{#1}$} \fi}
\newcommand{\tten}[1]{\ifmmode {\times 10^{#1}} \else {$\times 10^{#1}$} \fi}

\shorttitle{{\it SPEAR} Mission}
\shortauthors{EDELSTEIN et al.}

\begin{document}

\title{The Spectroscopy of Plasma Evolution from Astrophysical Radiation Mission}

\author{
J. Edelstein\altaffilmark{1}, 
K. W. Min\altaffilmark{2}, 
W. Han\altaffilmark{3}, 
E. J. Korpela\altaffilmark{1}, \\
 K. Nishikida\altaffilmark{1},
  B.Y. Welsh\altaffilmark{1},
   C. Heiles\altaffilmark{1}, 
     J. Adolfo\altaffilmark{1},
      M. Bowen\altaffilmark{1}, 
    W.M. Feuerstein\altaffilmark{1},
      K. McKee\altaffilmark{1}, \\
J.-T. Lim\altaffilmark{2},
 K. Ryu\altaffilmark{2}, 
 J.-H. Shinn\altaffilmark{2},
U.-W. Nam\altaffilmark{3}, 
J.-H. Park\altaffilmark{3}, 
I.-S. Yuk\altaffilmark{3}, 
H. Jin\altaffilmark{3}, 
K.I Seon\altaffilmark{3}, 
D.H. Lee\altaffilmark{3},
E. Sim\altaffilmark{4}
}

\affil{$^1$Space Sciences Laboratory, University of California, Berkeley, CA 94720} 
\affil{$^2$Korea Advanced Institute of Science and Technology, 305-701, Daejeon, Korea}
\affil{$^3$Korea Astronomy and Space Science Institute, 305-348, Daejeon, Korea}
\affil{$^3$Korea Aerospace Research Institute, 305-333, Daejeon, Korea}
\begin{abstract}

The Spectroscopy of Plasma Evolution from Astrophysical Radiation 
(or the Far-ultraviolet Imaging Spectrograph) instruments,
flown aboard the STSAT-1 satellite mission,
have provided the first large-area spectral mapping of 
the cosmic far ultraviolet  (FUV, $\lambda$ 900-1750)  background.
We observe diffuse radiation from
hot (10$^{4}$ -- 10$^{6}$ K) and ionized plasmas, 
molecular hydrogen, and dust scattered starlight.
These data provide for the unprecedented detection and
discovery of spectral emission from a variety of interstellar environments,
including the general medium, molecular clouds, supernova remnants, and super-bubbles.
We describe the mission and its data, present an overview of the diffuse
FUV sky's appearance and spectrum, and introduce the scientific findings
detailed later in this volume.

\end{abstract}


\section{Introduction}

The Spectroscopy of Plasma Evolution from Astrophysical Radiation instruments
(hereafter {\em SPEAR}, also known as the Far-ultraviolet Imaging Spectrograph or {\em FIMS})
have provided 
the first large-area spectral sky survey of cosmic far ultraviolet (FUV) radiation.
The FUV band ($\lambda$ 900-1750) includes important astrophysical diagnostics 
such as  strong atomic cooling lines from hot (T=10$^{4}$ K -- 10$^{6}$ K) 
and photoionized plasmas
as well as radiation from molecular hydrogen (H$_{2}$) fluorescent emission
and dust scattered starlight.
We describe the {\em SPEAR} mission and its science motivation,
introduce the data by presenting the spatial distribution of 
the FUV ($\lambda$ 1360 -1710) sky brightness 
and the spectra of the entire observed sky,
and provide a brief introduction to other $SPEAR$ science results 
described in the accompanying papers in this Volume.

{\em SPEAR} is the primary payload on the STSAT-1 (S-1) satellite
launched 2003 September 27.
The mission has thus far observed $\sim$80\% of the sky and 
conducted deep pointed observations toward numerous selected targets.
{\em SPEAR} contains dual imaging  spectrographs optimized for the measurement
 of diffuse FUV emission.
The spectrographs, referred to as the
Short wavelength band 'S'  ($\lambda\lambda $ 900 -- 1150, 4.0$^{\circ}$ x 4.6$\arcmin$ view)
and Long wavelength band 'L':  $\lambda\lambda$ 1350 -- 1750, 7.4$^{\circ}$ x 4.3$\arcmin$ view),
each have a spectral resolution of \ensuremath{\lambda /\Delta \lambda \sim 550} 
half-energy width
and an imaging resolution of 5$\arcmin$.
Each instrument uses a collecting mirror, a diffraction grating and an
open faced, photon counting micro-channel plate detector. 
The {\em SPEAR} instruments, their on-orbit performance, and the basic
processing of instrument data  are described in detail in the following paper
(Edelstein et al. \textit{ibid}.).

\section{Energetic Plasma in the ISM}

Supernovae and stellar winds produce shock-heated gas in the Galaxy. 
The activity powered by the energetic plasma shapes the structure 
of the Galaxy, effecting the distribution of metals and driving 
evolutionary phenomena such as star formation that depend on 
global morphology. 
On smaller scales, hot gas can power a variety of radiative, 
mechanical and chemical phenomena by forming
turbulent and evaporative structures, altering chemical abundance 
via dust ablation or vaporizing dust condensates, and by radiatively 
influencing the gas ionization balance. Candidate mechanisms 
(see \cite{mckee:95:ism} for an overview) by 
which interstellar plasmas cool include (1) energy exchange with surrounding 
media such as evaporation and conduction, (2) mechanical cooling 
such as adiabatic expansion, (3) direct emissive radiation, and 
(4) indirect radiation by way of heated dust-grain emission.  
Our understanding of shock-ISM interactions and the 
global structure of ionized gas in our Galaxy are far from complete.
Several models for the origin of this hot ionized gas have been proposed 
\citep{slavincox92,slavincox93,shapiro_benjamin91,Borkowski90,shelton98}
that make distinctive, verifiable predictions, yet none can 
be ruled out with current observations.

Very hot interstellar diffuse gas (T \texttt{>}= 10$^{6}$ K) was first mapped 
by {\it ROSAT}\, soft X-ray (SXR) background observations \citep{snowden98} 
which show pervasive regions of SXR gas at high latitudes,
conceivably  caused by hot gas injected into the Galactic halo,
although uncertainty remains regarding the state of this gas.
A component of this SXR emission has been attributed to local hot gas 
but charge-exchange between the Solar wind and the heliospheric environs
may confuse the observations by producing the same detected species 
\citep{lallement04_cx,margelin:charge_exchange:2004ApJ}.
\citet{sanders01_dxs} and \citet{Mccammon02}
have recorded SXR diffuse emission spectra that are inconsistent 
with predictions of standard models of hot ionized gas
but these data may also be contaminated by Solar charge exchange.
The \emph{CHIPS} mission \citep{chips05}
EUV observations
have established upper limits to local hot gas
that are an order of magnitude less than expected 
from the postulated local SXR emitting gas, unless
this plasma is extremely depleted.
(EUV observations are limited by interstellar absorption
to nearby regions, with N(HI) \texttt{<} 10$^{18.5}$ cm$^{-2}$.)

An overwhelming fraction ( $>85\%$) of the
radiative cooling power from hot, thin interstellar plasma 
in emitted in the FUV \citep{landini90}.
These FUV transitions, from
the must abundant atoms in prevalent ground states,
provide important diagnostics for 
both collisional and photoionized species.
FUV \emph{absorption} observations have revealed the hot ionized Galactic ISM. 
Measurements taken with \emph{IUE} and \emph{HST} have shown that
\SiIV, \CIV, and N V ions, characteristic of T = 10$^{4.5}$--10$^{6}$ K gas,
exist throughout the Galaxy with scale heights up to 4--5 kpc
\citep{sembach92_halo_ions}. 
Hotter gas, with T=10$^{5.2}$--10$^{6}$ K 
and indicated by O VI $\lambda\lambda$  1032 absorption, 
has been observed with \emph{Copernicus, Voyager, ORFEUS, HUT}
and now \emph{FUSE}
\citep{jenkins78a_o6, hurwitz95_orf_o6, davidsen93_hut, zsargo03_o6halo}. 
The O VI appears ubiquitous, although patchy 
in distribution, and perhaps extends to several kpc scale height
\citep{savage03_o6_halo}.
FUV absorption measurements, 
limited to sight lines with suitable background sources, 
cannot alone yield the physical state parameters of the hot plasma 
(e.g. n$_{e}$, T, pressure, filling factor)
and are not suited to the detection 
of lines from gas with a large velocity dispersion because placement of 
the adjacent stellar continuum is problematic. 

Detection of FUV interstellar \emph{emission} lines from the ISM
has proven to be difficult.
Measurements with sounding rocket experiments, short-lived 
orbital missions and small instruments on interplanetary missions 
have suffered from inadequate spectral or spatial resolution; 
an inability to carefully correct for noise sources such as intense 
geocoronal emission, bright stars and dust-scattered stellar 
continuum; or from integration periods insufficient to obtain sensitive results. 
(See \citet{bowyer91_araa} for a review of earlier work.)
Space observatories with FUV measurement capability 
(e.g., \textit{Copernicus, IUE, HUT, ORFEUS, GHRS, STIS, FUSE}) 
have been optimized for point source observations
and not for measuring faint diffuse spectra
over the large angular scales needed to characterize the Galactic plasma. 
Although the \textit{GALEX} mission  \citep{galex_miss_apjl}
was designed to observe large areas of the sky,
it has a limited capacity for mapping Galactic diffuse emission
and must exclude observations of regions including bright stars.
\textit{GALEX} records either a broad-band image including 
no spectral information, or an objective-dispersion image that 
obtains low resolution spectra of localized sources
but can confuse diffuse sources with angular extent.

Despite these difficulties, the detection of interstellar
\CIV \, and \OIIIb \, FUV line emission was reported \citep{martin90_c4} 
15 years ago (and not since, until now).
These lines were observed at intensities of several thousand 
LU\footnote{Line intensity units, ``LU'', are photons s$^{-1}$ cm$^{-2}$ sr$^{-1}$.
Continuum intensity units, ``CU'', are LU \AA$^{-1}$.}
toward four locations, 
from which the properties of an interstellar component were derived,
with n$_{e}$= 0.01 -- 0.02 cm$^{-3}$ 
and T = 10$^{4.7}$--10$^{5.3}$ K \citep{shull_slavin94}.
Emission from interstellar 
\OVI \, has been convincingly detected more recently,
with a doublet  intensity of \ensuremath{\sim}3000 - 5000 LU
toward about a dozen targets,
using long (20-200 ksec) \textit{FUSE} observations of 30'' fields 
\citep{dixon01_o6em,welsh02_o6em,shelton01_o6em,otte03_o6em}.  

The \emph{SPEAR} mission
was specifically designed to provide a spectral  imaging survey of diffuse
 FUV emission from the ISM.
We report the detection of such emission from most of the sky
(see Figure 1).
These data provide the first extensive
spectral observations of FUV cosmic diffuse emission.
We have found diffuse FUV emission lines
emanating from both atomic and molecular species
in a variety of interstellar environments.

\section{The SPEAR Mission, Operations \& Data}

\emph{SPEAR}, aboard the S-1 satellite, 
was injected into a 700 km sun-synchronous orbit
at 98.2$^{\circ}$ inclination
with an orbital period of 98.5 minute and
a $\sim$34 minute eclipse.
Observations are scheduled to begin
$\sim$360 s after eclipse entry
and end $\sim$300 s before eclipse exit.
About 10 daily orbits are scheduled for astronomy observations.
One or two orbits per day are used for down-looking observations 
of the northern night-side aurora.
The 3-axis controlled spacecraft platform can use a star tracker to achieve
5$\arcmin$ pointing knowledge,
pointed accuracy of $\sim$6$\arcmin$, and a stability of $\sim$12$\arcmin$.

Various pointing modes are used for survey, target and calibration observations.
Sky survey observations are performed by rotating a spacecraft axis such that
the \emph{SPEAR} field of view is swept, in a ``push broom'' fashion,
 perpendicular to its long field of view and
in a 180$^{\circ}$ great circle from the north to south ecliptic pole via the anti-Sun direction. 
Following the anti-Sun progression in this way over one year would cause a full viewing 
of the sky with a maximum of overlapped exposure at the ecliptic poles 
and a minimum of exposure at the  ecliptic plane.  
Calibration observations of stars or small ($\lesssim$10$^{\circ}$) fields are performed by
a ``back and forth'' spacecraft rotation that sweeps the field of view over a limited angle.
The calibration pointing mode,
 together with reports of the spacecraft roll rate and the
view-field's ground-measured width,
provide the most accurate 
positional data and exposure times for the observation of point sources.
Fixed inertial pointing toward specific targets can also be performed.
Pointings avoid the Sun by 45$^{\circ}$, the spacecraft velocity vector by 60$^{\circ}$ 
and are limited to zenith angles of  $<$110$^{\circ}$.

The SaTReC Mission Operations Center, Daejeon, Korea
is used to control the mission and receive data.
Data are decommutated into spacecraft-time marked attitude
and \emph{SPEAR}-time marked science and engineering information.
The information is passed through a photon reduction processing pipeline 
(Edelstein et al. {\it ibid})
that performs
(1) engineering selection of valid photon events,
(2) correction of detector electronic drift,
(3) transformation from detector to physical \ coordinates, 
(4) correction of detector distortions,
 and  (5) marking of photons with Universal time.
A mission data processing pipeline creates 
(1) a time history of attitude knowledge,
(2) a \emph{SPEAR} to spacecraft time association history,
(3) a time-associated sky exposure history
and 
(4) photon lists with time-associated attitude information.
The mission data products, in concordance with the photon data products,
can then be combined to produce fluxed spectra and spectral sky maps.
 
Each photon event is mapped to the sky 
by using a combination of the spacecraft attitude information,
the  instrument's bore-sight offset from the spacecraft axis,
and the angular position of the event on the detector.
Absolute bore-sight is determined by correlating reconstructed sky 
images with a field of known bright stars.
The spacecraft attitude knowledge, scheduled for report at 1-2 s intervals,
is determined by a star-tracker and a gyroscope (gyro).
The tracker updates are designed to
set the gyro knowledge to $\sim$2$\arcmin$ accuracy every 5 s.
The gyro knowledge drifts at 0.2$\arcmin$ s$^{-1}$.
The knowledge error,
derived from the drift rate and tracker update history,
is assigned by the pipe-line processing for each photon.
The sky location at times in-between spacecraft attitude reports
is computed using a spherical coordinate interpolation.

Sky exposure is derived according to valid exposure intervals (typically 60 s)
that are determined by examining 
operational and telemetry-interruption markers
rks 
interleaved with the photon data. Time-marked exposure records
are produced every 1.0 s within valid intervals for each 
5' increment within the spectrometers' field-viewing angles.  
Each exposure event can then be
mapped to a sky position in a similar fashion as for photons.
Exposure events include 
a weighting factor to account for fractional-second intervals 
that may occur at the end of valid periods, for partial coverage of a sky pixel,
or for other effects such as processing dead-time or angular vignetting 
(see Edelstein {\it ibid}). 
Time synchronization between the instrument and the spacecraft
is based upon an interpolation between
spacecraft time reports and
a correction that uses precise, synchronized 1.0 Hz
and 10 Hz timing marks which have been
interleaved with the instrument data stream.
The timing correction requires careful handling of clock interruptions
or erroneous telemetry. Photon events are marked to a precision of 0.1 Hz
with an estimated timing accuracy of 0.25 s.
Events with indeterminate time synchronization or shutter position,
usually due to corrupted and missing telemetry, are eliminated and 
correspond to a $\sim$2\% of the data.

\subsection{Mission Performance}

\emph{SPEAR's} observational performance depends on 
sky coverage,  sky exposure and mapping accuracy.
We report on the first year's  (2003 November -- 2004 November) observations.
The S-1 mission was designed for a 2-year mission, 
however spacecraft engineering problems may preclude further
science observations. 
During the first year,
2450 orbits of observations were recorded.
The S-1 attitude control system behavior
limits the quality of  attitude knowledge reconstruction.
We find that  the bore-sight offset is not constant on a per orbit basis
due to a variation in the reporting of attitude.
Furthermore, the star tracker updates are sometimes lost, 
resulting in an attitude error caused by gyro drift.
Finally, especially toward the end of the year, star tracker updates
or attitude knowledge reports are less frequently available.
About 40\% of our data suffer from some form of these problems.

The precision of attitude reconstruction from multiple, 
overlapping sky viewings
is affected by the attitude system's stability.
We have developed a systematic approach
that uses \emph{SPEAR} itself as a star tracker
to correct for bore-sight variability in survey and calibration sweeps.
This method has been applied to the observation of individual targets
and to $\sim$850 survey sweep orbits.
For observations that cycle sweeps over one field,
the attitude timing delay is varied until
L-band stellar images converge.
Absolute bore-sight is then determined by using a 2-d correlation 
of the image to bright stars listed in the TD-1 catalog \citep{td1}.
For survey or other single sweep observations, 
a similar correlation is used between TD-1 catalog objects 
and the L-band sweep image, recomputed for variable timing delays.
A single star's image can be reconstructed from multiple viewings
to an effective resolution of 
10$\arcmin$ FWHM with a 10$\arcmin$ positional accuracy.
Survey sweep positional corrections in the sweep track direction are 
found to be $\pm$30$\arcmin$ (2-$\sigma$)
and $\pm$8$\arcmin$ (3-$\sigma$) in the cross track direction.
Because the S-1 attitude follows a time sequenced command program,
\emph{SPEAR} can produce an image of a sweep-observed region
even when attitude knowledge is compromised or absent.
We anticipate that the attitude correction for a large fraction 
of the data with yet-uncorrected attitude problems can be improved.

\section{The Diffuse FUV Sky: Data Reduction}

We derive a spectral sky-map of diffuse emission from these data, 
from which we introduce the appearance of the FUV sky and its total spectrum.
Spectral sky maps were created by binning
photon and exposure events using the HEALPix tessellation scheme
\citep{Healpix}  with $\sim$15\arcmin\, pixels
and $\lambda$ bins for the L and S bands of 1.0\AA\, and 1.5\AA, respectively.
The accumulations, 
made only for times when the derived attitude error is $\leq$ 30$\arcmin$,
contains 1.2$\times$10$^{7}$ and 1.4$\times$10$^{7}$ photons
for the L and S bands, respectively.
The corresponding sky exposure map  for the L-band, 
shown in Figure 1,
covers $\sim$80\% of the sky and includes features such as
deep exposures ($>$10 ksec) toward calibration and pointed study fields,
and exposures of $>$500 s degree$^{-2}$ 
near the north ecliptic pole where many survey sweeps overlap.
Apparent are regions where no coverage exists due 
to the aforementioned attitude problems, to operational interruptions,
and to detector-protective shutdowns while observing 
FUV intense regions such as the Galactic plane.
The integrated exposure for the instantaneous field of view is 987 ks,
with an average of 65 s degree$^{-2}$
for observed sky regions.
About 15\% of these observations were taken 
using the 10\% shutter aperture.
The S-band exposure map is similar to the L-band map,
although coverage is reduced due to the smaller field of view.



In order to obtain true images of the cosmic FUV sky,
these data were subject to further reduction to account for 
artifacts, evident when simply dividing the photon map by the exposure map,
and to account for detector background noise contribution. 
We proceed with artifact removal and analysis of the {\em SPEAR} L-band map,
for 100\% shutter-position observations,
because of their superior coverage and sensitivity
in comparison to the S-band data.
The artifacts,
e.g. streaking in the direction of survey sweeps and localized over-intense regions,
are due to systematic errors in exposure determination
and to the recording of data during times of high airglow 
or radiation background contamination. 
Over-intense regions due to erroneous mapping were eliminated
by only including orbits 
that contain exposure records and photons 
that map to $>$95\% of the identical sky pixels.
Orbits having excessive airglow or particle induced detector noise
were identified by using the 
H Lyman-$\beta$ $\lambda$ 1027 and O I $\lambda$ 1356
count rates as airglow monitors.
Entire orbits with an average Lyman-$\beta$ rate exceeding
5 s$^{-1}$ were eliminated.
For certain regions showing residual ecliptic streaking
that have been observed repeatedly using successive sweeps,
entire orbits were rejected when their average monitor rate 
exceeded twice the dispersion of all the orbital rates about their median.
The L-band spectral map, after artifact removal, contains 
7.2$\times$10$^{6}$ photons and 810 ks  
of full-field exposure.

To obtain the \emph{diffuse} cosmic background,
we attempt to remove bright concentrated sources,
e.g. resolved stars, from the map.
Because the FUV diffuse intensity varies by orders of magnitude
across the sky,
stars must be identified as \emph{locally} intense pixels.
An L-band total intensity map,
integrated over $\lambda\lambda$ 1360 -- 1710,
was adaptively binned by sky area
to attain a statistical signal to noise (S/N) $\geq$45,
value for each bin.
In this scheme, the sky-bin size is increased 
along HEALPix pixel boundaries, 
i.e. for contiguous pixel groupings of size 4$^{n}$, 
until sufficient counts exist within the bin.
In this way, bright objects retain small angular dimensions
while faint regions are averaged to larger angular dimensions
with values of improved significance.
Bright sources were identified 
in $\sim$3\% of the pixels as having an
intensity that exceeds 2.5 times the median
value of the enveloping 4$^{\circ}\times$4$^{\circ}$ region
(256 HEALPix bins).
(We did not attempt to identify stars using point-spread function
detection schemes in this preliminary work because, 
at each point in the sky, the mapped point-spread function 
is a composite of the elliptical instrumental spread-function 
whose orientation differs for each sweep direction.)

The first Galactic map of total diffuse $\it SPEAR$ L-band FUV flux, 
shown in Figure 2, 
is obtained as an end product of the data reduction
by eliminating the locally intense pixels from
the starting continuum map, followed by 
adaptively binning the pixels to S/N $\geq$10.


\section {The FUV Sky Brightness \& Spectra}

The diffuse FUV L-band sky map (Figure 2) shows the largest intensity 
toward the Galactic plane and other regions where bright 
early type stars co-exist with significant columns of interstellar dust,
such as the obvious features of 
the Sco-Cen association and the Magellanic Clouds.
Thus, the cosmic FUV flux distribution is consistent with
what is is generally believed to be its dominant component,
starlight scattered by interstellar dust \citep{bowyer91_araa}, 
and is therefore similar to maps of reddening, N(HI) and H-$\alpha$
\citep{dust_map,skyview:HI_map,skyview:halpha_map}
because both of these maps trace in some way
either interstellar dust or the strong (local) FUV radiation fields that are
required to produce the dust-scattered FUV continuum.



Far UV spectra of the sky are obtained by spatially binning
the \emph{SPEAR} L-band spectral sky-map. 
The integrated {\it faint-sky} (bright-star subtracted) 
and {\it bright-sky} (star-only) FUV spectra are shown in Figure 3,
together with example spectra for the type of stars
expected to dominate the direct and scattered stellar FUV background
\citep{henry.isrf.2002}.
The \textit{bright-sky} spectrum has 
a $\lambda\lambda$ 1360 - 1710 median intensity of $\sim$30k CU.
The \textit{faint-sky} diffuse spectrum has a band median intensity of 2.3k CU
that includes contributions from a true cosmic signal, 
a combination of direct unresolved and interstellar dust-scattered starlight,
and residual instrumental background.

The spectral fitting model of Korpela et al. ({\it ibid}) is used
to identify and estimate the strength of both the astronomical and instrumental 
components within the {\it faint-sky} signal.
We find a median detector dark-noise contribution of 460 CU,
the geo-coronal emission line, OI $\lambda$ 1356,
and a number of prominent spectral features which we identify
as diffuse astrophysical emission lines from the ISM.
Strong emission lines from abundant interstellar species in this bandpass
can typically be identified with confidence 
because they are usually well separated from each other 
and from night-time geocoronal lines,
although features from unresolved stars or scattered starlight
and from interstellar H$_{2}$ do add some confusion to the spectrum.
Line identifications and intensities of prominent features are shown in Table 1, 
along with the intensities' statistical error (10 - 20 \%)
and line-fit modeling error.


Lines found that are likely to originate from the warm/hot ISM
include the pronounced \CIV $\lambda\lambda$ 1549 feature
and the \OIVb / \SiIV $\lambda\lambda$ 1400/1403 blend,
previously detected and tentatively identified, respectively, 
by \citet{martin90_c4} at similar intensities.
Interstellar diffuse lines, newly discovered with {\it SPEAR},
include the \SiIIs $\lambda$ 1533 and \AlII \,$\lambda$ 1672 resonance lines
that could originate from either neutral or warm ionized ISM.
We attribute the bright 1533 \AA \,  feature to the \SiIIs $\lambda$
fine structure ground state transition given that 
the \SiII \, $\lambda$ 1527 ground state transition
should be optically thick for typical interstellar conditions
and therefore undetectable.
The \AlII\, $\lambda$ 1671 feature may have been misidentified as
the  \OIIIb \, $\lambda\lambda$ 1665 doublet in prior 
lower resolution observations \citep{martin90_c4}.
Also discovered with {\it SPEAR} is diffuse 1657 \AA \, emission from \CI,  
a species that exists in the ISM despite photoionization by the
FUV background \citep{jenkins_tripp_CI}.

The FUV  H$_{2}$ fluorescence emission, 
previously detected by \citet{martin_hurwitz90_h2},
presents prominent and recognizable features in the \textit{faint-sky} spectra
\citep{black_vanD:1987},
particularly at $\lambda$ 1608 and  $\lambda$ 1580. 
(See the $SPEAR$ observations of \Htwo in Eridanus by Ryu et al. $ibid$,
for examples of the H$_{2}$ fluorescence spectral signature and band
identification.)
We identify the feature at 1639.1 \AA \, as \HeII\, $\lambda$ 1640 diffuse emission, 
although the fit line width,
50\% larger than other identified features,
indicates spectral contamination from H$_{2}$ or from \OI \, airglow,
although the night-time only observations mitigates
potential \OI \, contamination.
Some spectral features appear in 
both the \textit{bright-sky} and the \textit{faint-sky} spectra,
such as those at $\lambda$ 1485 and  $\lambda$ 1520,
which we tentatively attribute to residual stellar features
and some possible contribution from the H$_{2}$ fluorescence bands.
The total intensity estimated 
for all of the prominent interstellar  atomic spectral lines is $\sim$10k LU,
and for the distinct H$_{2}$ fluorescence bands is $\sim$9k LU.

\section{$SPEAR$ Results}

These FUV continuum and emission line data contain much
astrophysical information.
The continuum distribution depends upon
the interstellar dust and radiation field,
the \Htwo fluorescence reveals the molecular distribution and destruction rate,
and the FUV emission lines indicate the ionization equilibrium state 
and energetics of the ISM.
We shall explore these matters in future publications and present
$SPEAR$ spectral-line sky maps,
work which requires comprehensive analyses 
that is beyond the scope of this introductory letter.
We refer the reader to accompanying papers following in this volume,
summarized below, that use $SPEAR$ data and 
provide detailed discussions 
on the identification and importance of 
the FUV spectral components in different interstellar environments.
 
\subsection{The Diffuse ISM} 

Korpela et al ({\it ibid})
present $SPEAR$ S and L-band spectra of the 
North Ecliptic Pole (NEP) region 
($\beta$ = +30$^{\circ}$, N(HI) 2-8$\times$10$^{20}$ cm$^{-2}$),
a region that has no obvious associations with active interstellar regions
and therefore represents a canonical sight-line through the ISM.
A large number of diffuse FUV atomic emission lines are detected in the NEP,
representing the warm neutral to the hot ionized medium.
The atomic lines, ranging from 1500 to 6000 LU in intensity,
include those identified in the L-band all-sky spectrum previously presented,
as well as \OVI, \CIII, \NII \, in the S-band, and \NIVb \, in the L-band.
Presuming that the resonance lines are not dominated by scattering of the 
interstellar radiation field,
it is found that the high ionization potential lines cannot be fit 
by a collisional ionization equilibrium plasma model 
even though these lines, modeled on a per species basis,
have consistent emission measures, 0.001 to 0.005 cm$^{-6}$pc,
over the 10$^{4.5}$ to 10$^{5.5}$ K temperature range.
Therefore, it appears that photo-excitation,
non-equilibrium effects or abundance variations
are important in any explanation of the spectrum.
In addition, substantial H$_{2}$ fluorescence emission is detected 
in the NEP, despite the region's moderately low N(HI).

\subsection{FUV Hydrogen Fluorescence  \& Continuum} 

{\em SPEAR} observations show that interstellar FUV H$_{2}$ 
fluorescence is ubiquitous,
consistent with FUV absorption observations 
\citep{shull:h2.fuse.2000,h2:gillmon:hilat}
that find H$_{2}$ over large portions of the sky.
Lee et al. ({\it ibid}) observes H$_{2}$ fluorescence in the Taurus 
cloud's halo but not from the dense cloud core,
a fact attributed to the core's opacity excluding the 
FUV radiation needed to induce fluorescence.
Ryu et al. ({\it ibid}) finds H$_{2}$ fluorescence emission 
over a large region about the Ori-Eri super-bubble
and is able to elucidate the geometry of the Ori-Eri region 
by comparing the FUV fluorescence
with H$_{\alpha}$ and reddening maps and models 
an  H$_{2}$ excitation temperature of $\geq$1000 K, 
larger than generally found for molecular gas in the Galactic disk.
The H$_{2}$fluorescence is also found in deep {\it SPEAR} observations 
of the Ori-Eri super-bubble interface to the ambient ISM (Kregenow et al. {\it ibid}).

The $\it SPEAR$ observation of the Taurus cloud
(Lee et al, {\it ibid}) shows a counter-intuitive anti-correlation of FUV continuum 
with visual extinction and IR dust emission. 
The FUV continuum map of Taurus provides an optical transfer relation
over a wide range of depths that can be used to
quantify the dust and illumination properties -- 
the cloud core appears to block more distant FUV flux 
while the cloud halo scatters local flux toward the observer.

\subsection{Supernova Remnants and Superbubbles}

$\it SPEAR$ data provide unique spectral images 
of two nearby and well studied SNRs, Vela SNR (Nishikida et al {\it ibid}) 
and the Cygnus Loop SNR (Seon et al {\it ibid}).
Both authors find
that the spatial distribution of FUV emission cannot be
simply predicted using visible H$_{\alpha}$ or X-ray emission maps.
The remnants' emission line images, recorded in
C\,{\sc iii}, O\,{\sc iii}], C\,{\sc iv} and  O\,{\sc vi} far UV lines,
show where radiative shocks with
velocities ranging from \ensuremath{\sim}100 - 200 km s$^{-1}$ prevail.
The work directly verifies that the FUV emission lines are important to SNR cooling.
For Vela, the combined luminosity of strong FUV lines exceeds
the 1.0 -- 4.0 keV X-ray luminosity by an order of magnitude,
with a FUV to X-ray flux ratio that is a few times larger
than that for the Cygnus Loop.  The Vela enhancement is attributed 
in part to its higher covering factor of relatively
dense material that can give rise to FUV bright shocks.  

Kregenow et al. ({\it ibid}) reports on {\em SPEAR} spectra imaged
across the shell-wall boundary surrounding the Ori-Eri superbubble,
an X-ray emitting interstellar cavity that may have been created by SNR or stellar winds.
The FUV spectra are rich and include lines of similar species 
and intensities as found in the all-sky spectrum and in the NEP (Korpela et al. {\em ibid}).
Kregenow's study
discovers a distinct correspondence of \OVI, \CIV, and \SiIIs \, with the 
shell-wall interface traced in H$_{\alpha}$,
an unprecedented diagnostic for this type of evolved structure.
The work finds that the boundary emission is may be explianed
by either a quiescent thermal interface or a shocked region.

\section{Conclusion}

The first FUV spectral imaging survey of the large fractions of the sky
has been taken with the {\it SPEAR} mission. 
The resulting map of total FUV radiation shows that
diffuse FUV cosmic radiation is concentrated 
where both hot stars and scattering dust coexist,
such as in the Galactic plane, young stellar associations, and the Magellanic clouds.
The spectra of the total-sky contains prominent lines from ionized gas,
including the previously observed \CIV\, and \SiIV\, emission,
and from newly discovered \CI, \SiIIs, \AlII, \HeII \, emission,
as well as from H$_{2}$ fluorescence.
Similar interstellar FUV emission lines 
are found in the general ISM, as sampled at the NEP,
and toward the Eri-Ori superbubble and its interface to the ISM.
Intense FUV emission lines from shocks are visible in nearby SNR.
Diffuse H$_{2}$ fluorescence is ubiquitous across observed Galactic environments.
Much work remains to elaborate upon the character of the FUV sky 
and to analyze the physical conditions of detailed objects.

\acknowledgments
\emph{SPEAR / FIMS} 
is a joint project of KASI \& KAIST (Korea) and U.C., Berkeley (USA),
funded by the Korea MOST and NASA Grant NAG5-5355. 
We thank the team for their remarkable effort to create the mission
and families and friends for their love and support.
Thanks Phil.

\bibliographystyle{apj}   




\pagebreak
\clearpage
\newcommand{\tableone}{
\begin{deluxetable}{lccrcc}
\tablecaption{Identification of {\it faint-sky} line features
\label{him_lines}}
\tablehead{
\colhead{Species} & \colhead{$\lambda$} & \colhead{$\lambda_{fit}$} 
& \colhead{$I_\lambda$} &  \colhead{Statistical} & \colhead{Model} \\
\colhead{\ } & \colhead{\ }& \colhead{\ }  & \colhead{\ } 
& \colhead{Error} & \colhead{Uncertainty} \\
\colhead{\ }  & \colhead{\angstrom} & \colhead{\angstrom} 
& \colhead{LU}\footnote{LU=\lu} & \colhead{LU} & \colhead{LU} 
}
\startdata
\SiIV \footnote{blended with OIV] 1400}	&	 1403 	&	1403.1	&	736	& 162 & 281\\
\SiIIs 	&	1533	&	1532.0	&	1909	& 201 & 275 \\
\CIV	 	&	1550	&	1548.7	&	2636	&	211& 341  \\
H$_{2}$	&	1608	&	1606.5	&	1832	&	262 & 385 \\
\HeII	&	1640	&	1639.1	&	3577	&	378 & 1177\\
\CI	&	1657	&	1655.7 &	1316 	&	276 & 487\\
\AlII	&	1670	&	1669.7	&	1474	&	280	 & 226 \\

\enddata
\end{deluxetable}
}
\tableone


\pagebreak
\clearpage
\begin{figure}[ht]
\plotone{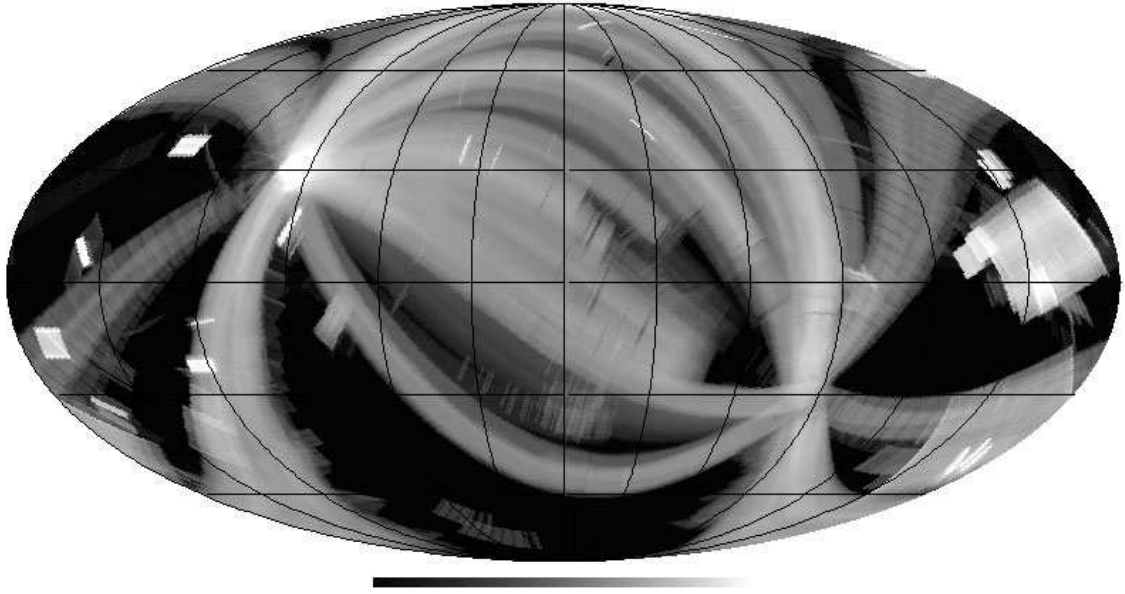}
\caption{
Sky exposure for 
the \emph{SPEAR} L-band observations.
The map is made with
${0.5^{\circ}}$ pixels,
a histogram equalized intensity scale with a maximum of 500 s degree$^{-2}$,
Galactic Aitoff coordinates centered at $l,b$ = 0$^{\circ}$,0$^{\circ}$
with longitude increasing toward the left,
and is shown with latitude and longitude lines on a 30$^{\circ}$ grid.
}
\label{exposure}
\end{figure}


\pagebreak
\clearpage
\begin{figure}[ht]
\plotone{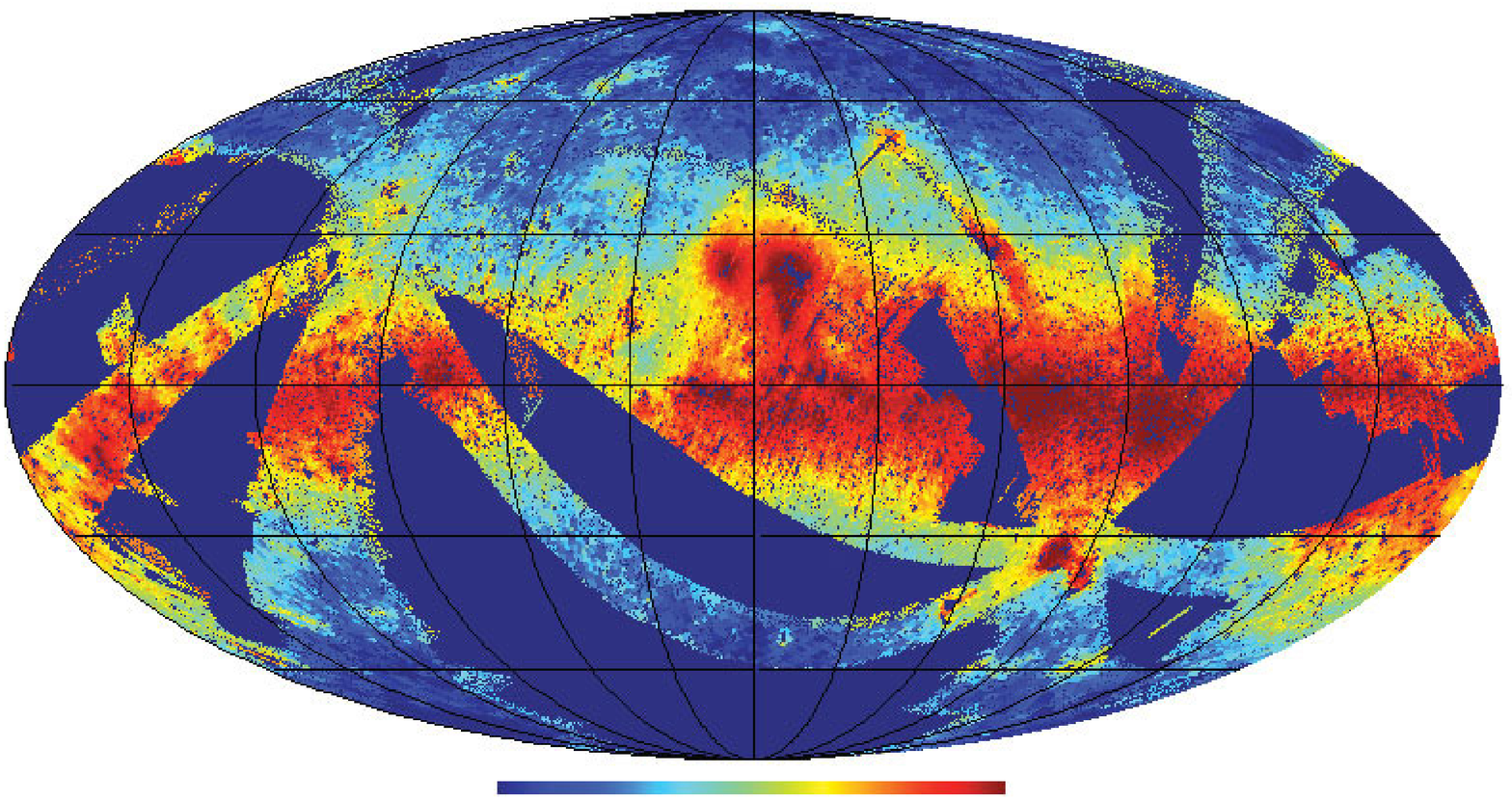} 
{\caption
Total diffuse intensity map of the sky for the \emph{SPEAR} L-band
($\lambda\lambda$ 1360 - 1730) observations,
after removal of locally intense pixels (stars).
The map, in the same coordinates scheme as Figure 1,
has a histogram equalized logarithmic intensity scale
with a maximum of 20k CU,
and is adaptively binned by sky area to a S/N $\geq$10.
Evident features include the Galactic Plane, the Sco-Cen association
(e.g. zeta-Oph at $l,b$ = 6$^{\circ}$, 24$^{\circ}$),
and the LMC at $l,b$ = 280$^{\circ}$, -32$^{\circ}$.
}
\label{skymap}
\end{figure}



 \pagebreak
 \clearpage
\begin{figure}[ht]
\plotone{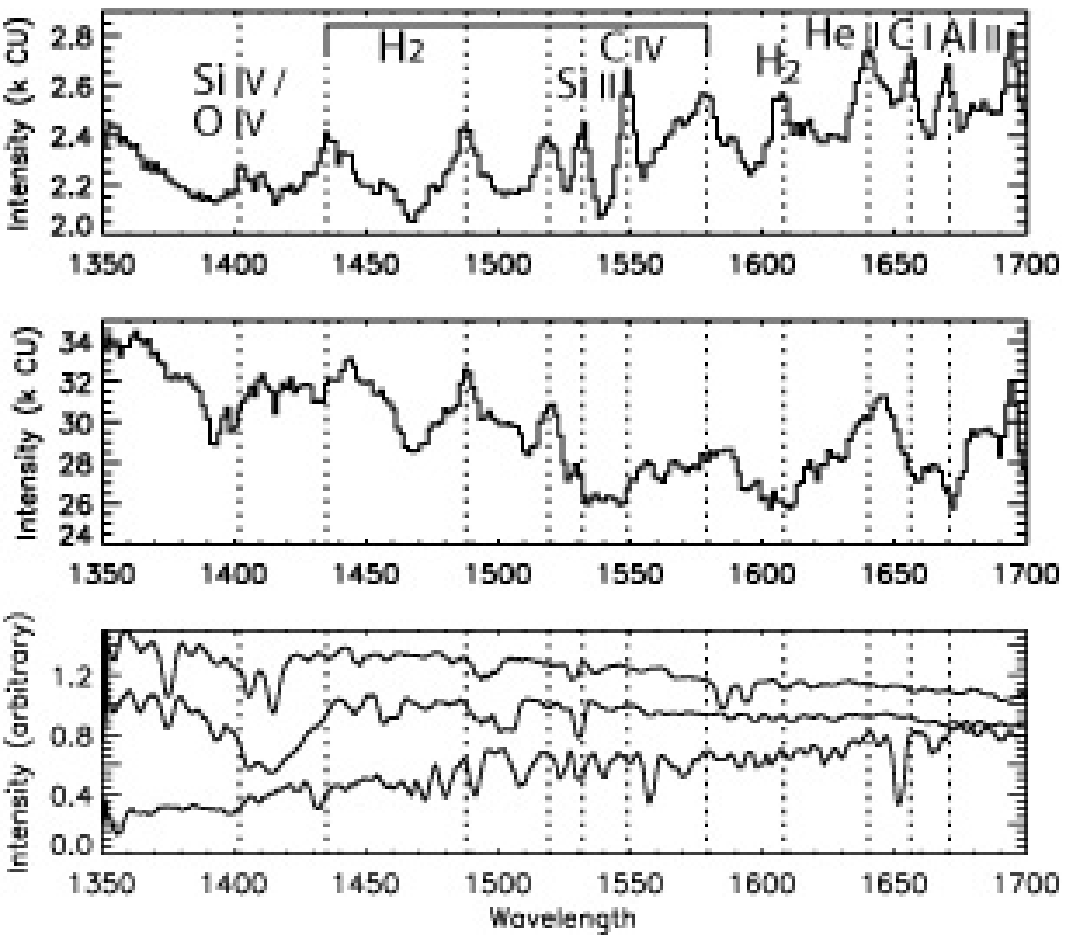}
\caption{ 
The  L-band spectra 
{\it (top)} from the {\it faint-sky} and
{\it (middle)} from the {\it bright-sky}, plotted with false zeros.
 {\it (bottom)} {\it IUE} measured stellar spectral for B3V, B5V and B8V stars in arbitrary units.
Dashed lines mark prominent  {\it faint-sky} spectral features.
}
\label{spectra}
\end{figure}


\end{document}